# Up to 70 THz bandwidth from an implanted Ge photoconductive antenna excited by a femtosecond Er:fibre laser

Abhishek Singh[1], Alexej Pashkin[1*], Stephan Winnerl[1], Malte Welsch[1,2], Cornelius Beckh[3], Philipp Sulzer[3], Alfred Leitenstorfer[3], Manfred Helm[1,2], and Harald Schneider[1]

[1]Institute of Ion Beam Physics and Materials Research, Helmholtz-Zentrum Dresden-Rossendorf, 01328 Dresden, Germany
[2]Cfaed and Institute of Applied Physics, TU Dresden, 01062 Dresden, Germany
[3]Department of Physics and Center for Applied Photonics, University of Konstanz, 78457 Konstanz, Germany

**Phase-stable electromagnetic pulses in the THz frequency range offer several unique capabilities in time-resolved spectroscopy. However, the diversity of their application is limited by the covered spectral bandwidth. In particular, the upper frequency limit of photoconductive emitters - the most widespread technique in THz spectroscopy – reaches only up to 7 THz in the regular transmission mode due to the absorption by infrared-active optical phonons. Here, we present ultra-broadband (extending up to 70 THz) THz emission from an Au implanted Ge emitter which is compatible with modelocked fibre lasers operating at 1.1 and 1.55 μm wavelengths with pulse repetition rates of 10 and 20 MHz, respectively. This result opens a perspective for the development of compact THz photonic devices operating up to multi-THz frequencies which are compatible with Si CMOS technology.**

### Introduction

THz time-domain spectroscopy using broadband THz pulses has emerged as a powerful tool for probing low-energy excitations in condensed matter at the meV energy scale.[1-3] The spectrum of potential applications depends on the spectral bandwidth, signal-to-noise ratio and data acquisition speed available. In general, the techniques for THz generation and detection exploit either photoconductivity or optical nonlinearity.[4,5] Photoconductive techniques for THz emission and detection are widely used due to their simplicity, compactness and the possibility of direct coupling to fibre optics. Although THz emission from photoconductivity was first demonstrated using Si,[4,6,7] the majority of photoconductive antennas these days are based on GaAs or InGaAs (in case of telecom wavelength) due to the high carrier mobility in these materials and well-established schemes for reducing the carrier lifetime.[8] Optical rectification techniques rely mostly also on polar non-centrosymmetric materials with strong second-order optical nonlinearity such as ZnTe, GaP, GaSe or *DSTMS*.[9] The polar nature of all these materials renders their optical phonons strongly IR-active leading to reststrahlen bands in the region between 5 and 10 THz. Due to this fact, the spectral bandwidth of many THz emitters is limited to below 7 THz in the regular transmission mode. In particular, for InGaAs-based photoconductive emitters excited at 1.55 μm wavelength gapless THz spectra up to 6.5 THz have been demonstrated.[10] Thin electro-optic crystals of GaSe and DAST have shown THz emission extending up to more than 100 THz towards the higher frequency end, but the THz intensity near their phonon frequencies is strongly suppressed.[11-14] Even in the reflection geometry available with photoconductive emitters, strong absorption and emission by polar TO and LO phonons, respectively, hinders their application for spectroscopy around the resonance frequencies.[15,16]

To fulfil the demand of a gapless ultra-broadband spectrum, novel techniques such as two-color air plasma[17] and spintronic THz emission[18] have been introduced. THz emission from air plasma demonstrates a bandwidth of more than 100 THz, but this technique requires high pump-pulse energies of several 100 µJ or higher that can be achieved only by rather complex and expensive laser amplifiers.[17,19] Spintronic emitters have shown great potential as a gapless broadband emitter reaching up to 30 THz bandwidth being compatible with nJ laser pulses from conventional femtosecond oscillators.[18] Recently their scalability for generation of higher THz fields has been also demonstrated.[20] A similar study by Wu et. al. demonstrates an efficient operation of such THz emitter driven by pump power as low as 0.15 mW.[21] Nevertheless, the THz generation using photoconductive antennas remains important for many applications due to the direct control of the THz field strength and polarity by applied bias voltage. Moreover, specially designed electrode geometries enable generation of radial or azimuthal THz polarizations[22] as well as fully controllable angle of the linear polarization.[23,24] However, until recently the bandwidth coverage of photoconductive emitters has been limited by the above-mentioned factors.

A breakthrough in generation of a broadband THz spectrum beyond the reststrahlen band of III-V semiconductors has been achieved recently by using a Ge-based photoconductive dipole antenna based on pure Ge.[25] This semiconductor has a direct interband absorption above 0.8 eV – very close to its indirect bandgap at 0.66 eV. The effective electron mass in the centre of the Brillouin zone of Ge is fairly small leading to a strong acceleration of photogenerated electrons and, correspondingly, to efficient THz emission. This property gives Ge a clear advantage over Si in applications for photoconductive THz devices. Moreover, the relatively small bandgap of Ge enables pumping with compact fibre lasers. Finally, The absence of polar phonons in Ge enabled generation of a gapless THz spectrum spreading up to 13 THz. Ge is known to be compatible with Si CMOS technology[26] and, thus, it is attractive for integrated on-chip THz solutions for THz signal processing.[27,28]

The absence of polar phonons in Ge enabled generation of a gapless THz spectrum spreading up to 13 THz and it has been demonstrated that the bandwidth of the Ge-based THz emitter is limited only by the duration of the excitation and detection laser pulses and, therefore, can be potentially extended to much higher frequencies.[25] However, an ultimate performance can hardly be reached for intrinsic Ge due to the relatively long carrier lifetime of several µs caused by the indirect character of the bandgap. The repetition rate of the driving laser must be low enough (250 kHz and less) to ensure full recombination of the carriers between the pulses limiting the choice to complex and expensive regenerative laser amplifiers for which sub-30 fs pulse duration is cumbersome to achieve and the full pulse energy usually cannot be exploited due to saturation of the THz emission by screening effects.

**Results**

In order to harness the full potential of Ge as a material for photoconductive THz emitters we reduce the carrier lifetime down to sub-nanosecond level by introducing deep traps via Au implantation. Although a shorter carrier lifetime is not an essential requirement for the broadband THz emission, sub-ns lifetime ensures reliable operation of Ge:Au THz emitters at repetition rates up to few 100 MHz covering specifications of most contemporary femtosecond oscillators. It is known that Au in Ge forms deep acceptor levels within the bandgap which possess large capture cross sections and drastically reduce carrier lifetime already for very low doping concentrations.[29,30] Ge substrates were implanted with Au ions with energy of 330 keV and doses of $5\times10^{13}$ ions/cm$^2$ and $2\times10^{13}$ ions/cm$^2$ followed by annealing at 900 ℃ for several hours in order to ensure a low homogeneous concentration of gold impurities near the surface of the Ge wafer. After annealing, Au ions diffuse hundreds of µm deep inside the Ge substrate resulting in a suitable doping density of approximately $10^{15}$ cm$^{-3}$.[31]

We have estimated the carrier lifetime in implanted Ge wafers using optical pump / THz probe spectroscopy. Fig. 1 shows the photoinduced change in THz transmission ΔT which is approximately proportional to the density of free charge carriers. The comparison between pure and Au-doped Ge clearly demonstrates a dramatic reduction in the recombination time. The pure Ge sample shows a step-like increase of carrier density with a minor drop in the following 1.5 ns. Moreover, there is a strong non-zero response at negative delay times indicating a high density of carriers accumulated in the sample from the preceding pump pulse arriving 4 μs earlier. In stark contrast, Ge:Au samples show a strong decay within

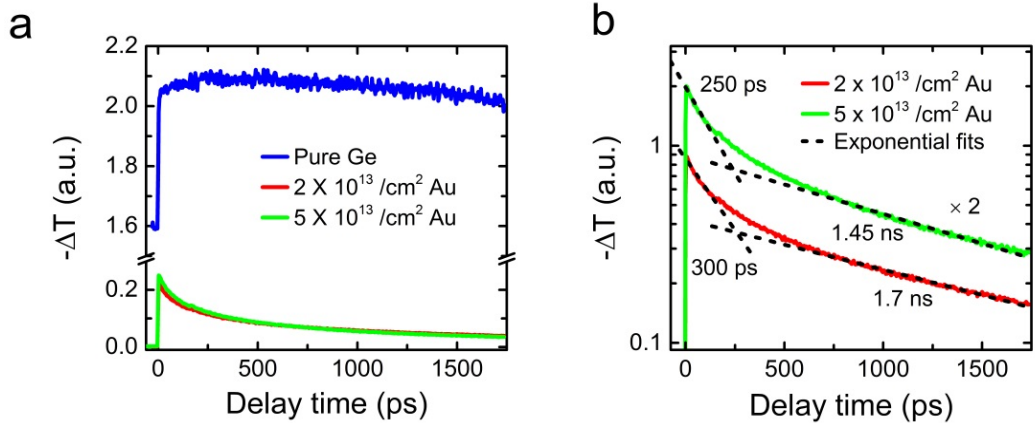

**Fig. 1 Carrier recombination in Ge measured by optical pump / THz probe spectroscopy. a,** Pump-induced change in THz transmission as a function of the pump-probe delay time for pure and Au-implanted Ge. Pump fluence is 12.5 μJ/cm², wavelength 800 nm. **b,** Bi-exponential fits of the decay dynamics for Ge:Au samples. The curves for 5 × 10¹³ ions/cm² dose are vertically shifted for clarity (multiplied by 2).

1 ns after photoexcitation and a negligible offset at negative delay times. The recombination dynamics can be described using a bi-exponential decay as shown in Fig 1b. The faster decay time of ≈300 ps can be attributed to a surface recombination and the slower nanosecond decay to the trap-assisted recombination in the volume of the Ge:Au sample. Both implanted Ge substrates exhibit a carrier lifetime of less than 2 ns which is about 3 orders of magnitude shorter than the typical carrier lifetime in pure germanium.

Bowtie electrode structures with 10 μm gap and 30 μm length of each electrode, were fabricated on two implanted Ge:Au substrates using the same fabrication process and the bowtie electrode geometry as our previous work.[25] The schematic diagram showing its operation is presented in Fig. 2(a). The near-infrared pump beam is focused onto the 10-μm-gap between the electrodes. Photoexcited charge carriers are accelerated by the applied bias field producing a transient current burst. The THz beam emitted by this current in forward direction is collimated and refocused on an electro-optic crystal for field-resolved detection. The short carrier lifetime enables us to operate the Ge emitter at the repetition rate of tens of MHz using a femtosecond fibre laser system.[31]

First, we test the THz emission induced by 11-fs-short pulses with a central wavelength around 1100 nm (spectrum spanning from 900 to 1250 nm) and an energy of 7 nJ (at the repetition rate of 10 MHz). The emitter was fabricated on Ge:Au with the implantation dose of 2 × 10¹³ ions/cm². The generated THz transient is detected by electro-optic sampling using 8.42-fs-short probe pulses in a (110) ZnTe crystal with a thickness of 14.3 μm. Figs. 2b & 2c show the recorded THz pulse in the time domain and its Fourier spectrum, respectively. The obtained spectrum spans from the lowest detectable frequency up to 70 THz demonstrating unprecedented bandwidth for a photoconductive antenna with a gapless spectrum. The small dip in the spectrum around 5 THz is caused by the response function of the ZnTe detector and the true THz emission spectrum is gapless as it has been confirmed in previous work.[25] The estimated peak electric field of the focused THz pulse is ~ 0.8 kV/cm and the signal-to-noise ratio is ~300.

We have modelled the detected THz spectrum using the intensity and phase spectra of the pump and probe near-infrared pulses.[31] The emission from the biased Ge:Au is calculated by solving the equation for the pump pulse propagation in Ge which takes into account the absorption and the dispersion of the refractive index. In this way, the temporal and depth profile of the photoinduced carrier concentration has been calculated. The total emitted THz field is estimated as a sum of the emission by currents across the full depth of the emitter. The resulting spectrum is multiplied by the detector response function (DRF) of the ZnTe detector which takes into account the group velocity dispersion across the very large bandwidth of the probe pulse. Further details are given in the Supplementary Material.[31] The effect of diffraction limited focusing of the THz beam on the detector is also taken into account. The result shown in Fig. 2d agrees well with the experimentally measured spectrum demonstrating that roll-off at higher frequency is mainly due to widths of the pump and probe pulses, and roll-off towards lower frequency is due to diffraction limited focusing of THz on the detector crystal. Thus, we observe a gapless spectrum extending up to 70 THz (wavelength of 4.3 μm).

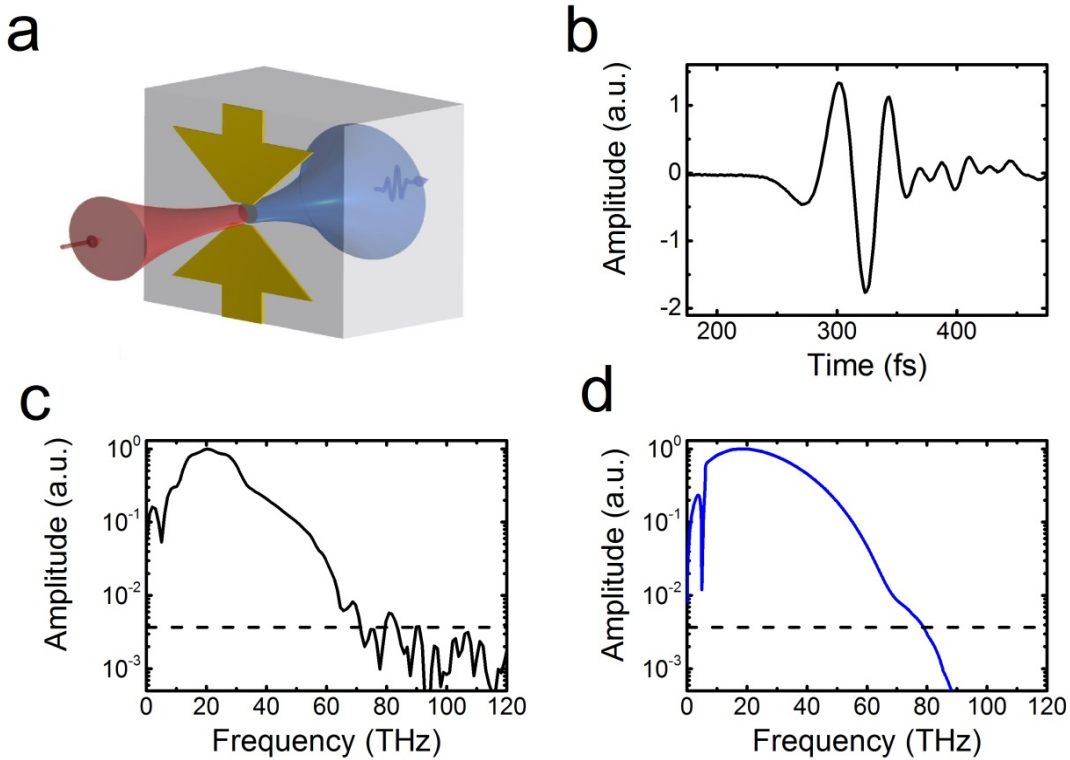

**Fig. 2 Ultrabroadband THz emission from Ge:Au antenna pumped at 1100 nm. a,** Schematic diagram of the Ge photoconductive emitter with bowtie-like electrodes. Pump and THz beams are shown in red and blue, respectively. **b,** Recorded THz pulse for pumping with 11 fs pulses with a wavelength centred at 1100 nm. **c,** Fourier transform of the recorded THz pulse. **d,** Simulated THz spectrum calculated as a product of the THz field emitted by the photoinduced current in the Ge emitter and the ZnTe detector response function. Dashed lines show the experimental noise floor.

The performance of the implanted Ge emitter is studied at varying pump powers and applied bias. Figs. 3a and 3b show the observed variation in the peak-to-peak electric field amplitude of the emitted THz pulse at different pump powers and applied bias, respectively. Usually the peak-to-peak electric field amplitude of photoconductive emitters increases linearly with pump power until the saturation sets in due to screening of the applied electric field. The implanted Ge emitter also shows similar behaviour with pump power as shown in Fig. 3a. The THz field scales almost linearly for pump pulse energy up to 3 nJ (30 mW power at 10 MHz) showing saturation beyond that limit. Similarly, the THz field is also expected to scale linearly with the applied bias field on the emitter electrodes. Fig. 3b demonstrates almost linear dependence with signal vanishing at zero bias, thus confirming that the THz emission is produced solely by the photoinduced current in the Ge antenna.

In the following we demonstrate the compatibility of the Ge:Au photoconductive antenna with the conventional telecom C-band covered by ubiquitous Er-doped femtosecond fibre lasers. The absorption edge of Ge at 0.66 eV ($\lambda \approx 1876$ nm) is well below the telecom band. However, the absorption is rather weak due to the indirect character of the bandgap and increases abruptly only for photon energies above 0.8 eV ($\lambda \approx 1550$ nm) corresponding to direct interband transitions near the Γ-point of the Brillouin zone.[32]

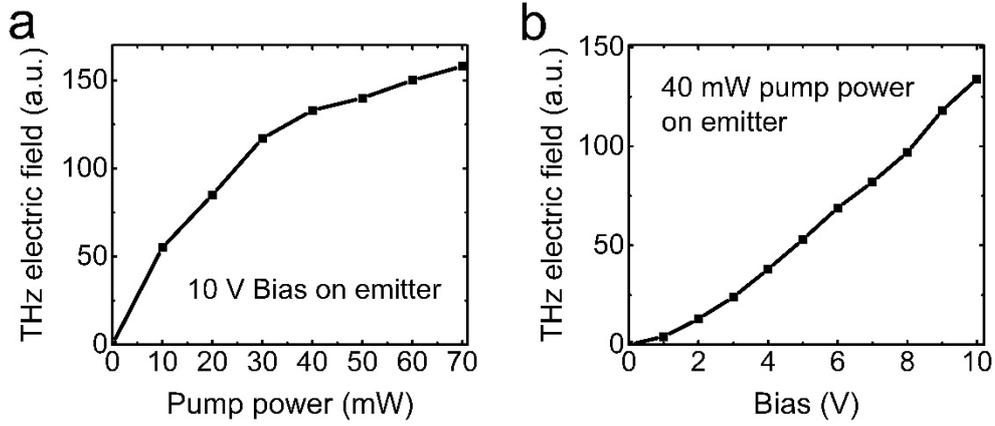

**Fig. 3 Peak-to-peak electric field of the THz pulse. a** THz peak-to-peak electric field variation at constant bias of 10 V and varying pump powers. At high pump-pulse energy the separation of photogenerated electron-hole pairs causes the screening of the applied d.c. field on the emitter. Hence the emitter's efficiency gets affected causing the saturation of the emitted THz field **b** THz peak-to-peak electric field variation at constant pump power of 40 mW and varying applied bias on the emitter electrodes.

The onset of the strong absorption in Ge coincides with the telecom wavelength of 1550 nm and, thus, should enable a broadband emission from the Ge-based THz emitter.

The emitter fabricated on Ge:Au with the implantation dose of $5 \times 10^{13}$ ions/cm$^2$ was pumped with 12 fs pulses with a central wavelength around 1550 nm and the energy of 3.5 nJ at the repetition rate of 20 MHz. The spectrum of the pump pulse is shown in Fig. 4a together with the absorption coefficient of Ge demonstrating that the part of the pump pulse with photon energies below 0.8 eV should contribute much less to the generation of the transient photocurrent than its high energy part. Thus, even though the initial pulse duration and the energy are comparable to the previous case of the pumping at 1100 nm, the expected THz bandwidth should be lower than 70 THz.

The emitted THz pulses are recorded with an 18 μm thick ZnTe detector crystal by electro-optic sampling using 5.8 fs pulses with a central wavelength of 1200 nm.[31] The photoconductive antenna is pumped at the repetition rate of 20 MHz under applied d.c. bias of 10 V. Figs. 4b and 4c show the recorded THz waveform and its Fourier transform, respectively. The THz spectrum spans up to 50 THz demonstrating that the Ge-based photoconductive antenna is capable for broadband THz emission when pumped at telecom wavelength. As anticipated, the bandwidth and the dynamic range of the THz waveform is lower than for pumping at 1100 nm due to non-uniform absorption of different parts of the excitation spectrum. The estimated peak electric field is ~ 0.12 kV/cm and the signal-to-noise ratio is ~65. Furthermore, the pulse shape differs from the typical single-cycle THz waveform that is observed for pumping at 1100 nm (see Fig. 2b) and includes an additional multi-cycle THz component with a frequency of approximately 13 THz. The phase of this narrowband feature is nearly opposite to the broadband THz pulse at low frequencies leading to the minimum in the detected spectrum below the 13 THz peak due to destructive interference.

## Discussion

Although a similar narrowband emission by LO phonons in polar semiconductors is well known and understood,[15,33] such a component is not expected in non-polar Ge. In order to verify whether this emission can be related to indirect interband absorption away from the surface of the photoconductive antenna we have performed a simulation of the recorded THz pulse using the same approach as for the 1100 nm pumping. The time-dependent current across the full depth of the Ge:Au wafer has been calculated by numerically solving the pulse propagation equation.[31] The resulting THz spectrum depicted in Fig. 4d reasonably describes the decrease of the bandwidth to 50 THz due to the dominating role played by the carriers photoexcited via the direct interband transitions for photon energies above 0.8 eV. In fact, our simulation shows a negligible contribution of the spectral components below 0.8 eV.[31] However, besides that the THz emission is expected to be similar to the case of the pumping at 1100 nm with nearly single-cycle THz waveform. Thus, the narrowband emission at 13 THz cannot be attributed to a standard transient photocurrent in our structure. A more exotic mechanism such as coherent polarizations due to a simultaneous generation of heavy-hole–light-hole wavepackets as recently reported in GaAs[34] may be considered. However, this discussion is beyond the scope of the present work and also requires a larger amount of experimental data.

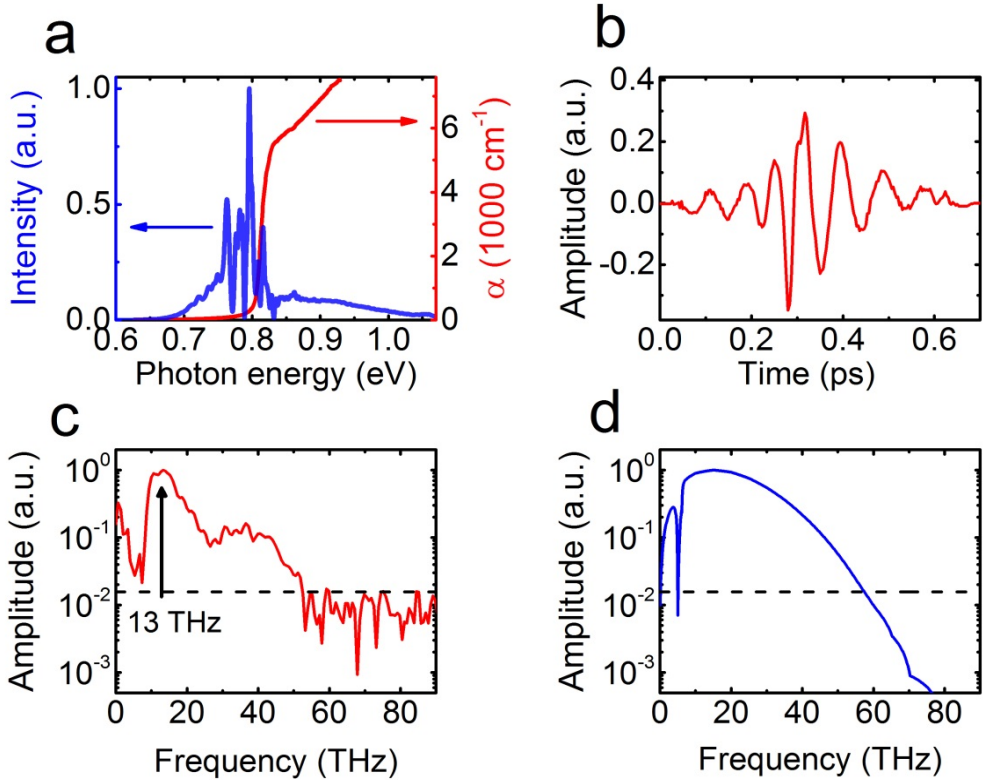

**Fig. 4 Broadband THz emission from Ge:Au antenna pumped at 1550 nm. a,** Spectrum of the pump pulse (blue) as compared to the absorption coefficient of Ge (red) taken from Ref. 29. **b,** Recorded THz pulse for the pumping with 12 fs laser pulses centred around 1550 nm. **c,** Fourier transform of the recorded pulse. The arrow marks the contribution of the narrowband part of the pulse at 13 THz. **d,** Simulated THz spectrum. Only the broadband single-cycle pulse stems from the photocurrent. Dashed lines show the experimental noise floor.

In conclusion, we have demonstrated a photoconductive THz emitter fabricated on Au-implanted Ge, which is capable of emitting a gapless spectrum with unprecedented bandwidth reaching 70 THz. The tested devices are fully compatible with femtosecond fibre lasers and demonstrate ultrabroadband THz emission under the pumping at either 1100 nm or 1550 nm. Thus, Ge-based THz emitters may be used for the generation of a gapless spectrum also in combination with standard Er-doped femtosecond fibre lasers at frequencies as high as 76 MHz. The demonstrated bandwidth is almost one order of magnitude higher as compared to that of existing state-of–the-art photoconductive THz emitters fabricated on GaAs or InGaAs. Consequently, Ge-based THz devices can revolutionize THz technology due to its ultrabroad spectral bandwidth coverage and its potential compatibility with Si CMOS technology.

## Materials and Methods

### Ge implantation

Au ions with 330 keV energy and with doses of $2\times10^{13}$ ions/cm$^2$ and $5\times10^{13}$ ions/cm$^2$, respectively, were implanted into two nominally undoped (100) Ge substrates. Simulation using the SRIM software shows the initial implantation depth of approximately 150 nm (supplementary Fig. S1). To distribute the Au ions uniformly over a length scale larger than the penetration depth of the pump light and to recover the lattice damage after ion irradiation, samples were annealed in vacuum at 900 °C for 3 hours (sample with dose $5\times10^{13}$ ions/cm$^2$) and 10 hours (sample with dose $2\times10^{13}$ ions/cm$^2$), respectively Annealing is expected to cause Au diffusion for more than 100 µm deep into the Ge wafers.[31] After the processing, the samples' surfaces were polished to make them smooth enough for lithographic processing.

### Optical pump / THz probe measurements

The measurements of the carrier lifetime in Ge:Au were performed using a THz setup equipped with a GaAs-based large-area photoconductive emitter and ZnTe electro-optic detector covering frequencies up to 3 THz. The system is driven by a Ti:Sa amplifier laser operating at the repetition rate of 250 kHz and the wavelength of ~ 800 nm. 50 and 200 mW pump power is used to pump the ≈3 mm diameter area resulting in the fluence of 12.5 and 50 µJ/cm$^2$ for the excitation of pure and implanted Ge substrates, respectively. The THz pulse is focused to about 1 mm size on the same spot ensuring homogeneous pumping conditions. The change in the THz transmission is measured using lock-in detection at different pump-probe delay times. The data shown in Fig. 1 are recorded at the peak of the THz probe pulse and normalized with respect to the pump fluence in order to provide a direct comparison between the different curves.

### Emitter fabrication

Bowtie-like electrodes were fabricated on the two implanted Ge substrates using standard electron beam lithography. Two layers of 5 nm Ti and 45 nm Au were deposited one after another and the bowtie geometry was formed by a lift-off process.

### THz emitter characterization at 1550 nm and 1100 nm pumping

The THz setup is based on ultrabroadband Er:fibre laser technology.[35] The repetition rate of the electro-optic sampling pulses with the wavelength of 1100 and 1550 nm are 20 and 40 MHz, respectively. The pump pulses with 1100 nm or 1550 nm wavelength are modulated at 10 and 20 MHz, respectively – the half of the oscillator frequency enabling excellent sensitivity of the system at the shot-noise limit using a lock-in detection technique. Further details can be found in additional materials of recent publications that utilized the same fibre laser systems.[36,37]

A 18-µm-thick ZnTe crystal is used for the electro-optic sampling of the emitted THz pulses when the emitter was pumped with 1550 nm wavelength. 14.3-µm-thick ZnTe and 18.4-µm-thick GaSe are used to characterize the emitter pumped with 1100 nm wavelength. The studied photoconductive emitters were biased a by static voltage, since the pump pulses are already modulated at half the probe repetition rate.

**Acknowledgements**

Support by R. Böttger and the Ion Beam Center (IBC) at HZDR is gratefully acknowledged. The authors would like to thank W. Skorupa for his valuable suggestions as well as I. Skorupa and U. Lucchesi for technical assistance.

**Authors' contributions**

A.S. and A.P. conceived the experiments and performed the doping of the Ge wafers. A.S. measured the carrier lifetime, fabricated the antenna structure and obtained the THz signal using a Ti:Sa laser. C.B., P.S. and A.L. constructed the THz setups based femtosecond Er:fibre lasers. C.B., P.S. and A.S. carried out the THz measurements using the fibre laser systems. A.S., M.W. and A.P. performed numerical modelling of the THz signals. The paper was drafted by A.S. and A.P. All authors contributed to discussing the results and writing the paper.

# Up to 70 THz bandwidth from an implanted Ge photoconductive antenna excited by a femtosecond Er:fibre laser

Abhishek Singh[1], Alexej Pashkin[1], Stephan Winnerl[1], Malte Welsch[1,2], Cornelius Beckh[3], Philipp Sulzer[3], Alfred Leitenstorfer[3], Manfred Helm[1,2], and Harald Schneider[1]

[1]Institute of Ion Beam Physics and Materials Research, Helmholtz-Zentrum Dresden-Rossendorf, 01328 Dresden, Germany
[2]Cfaed and Institute of Applied Physics, TU Dresden, 01062 Dresden, Germany
[3]Department of Physics and Center for Applied Photonics, University of Konstanz, 78457 Konstanz, Germany

**Au implantation and diffusion due to annealing**

The Au ion distribution in Ge wafer after the 330 keV ion energy implantation is simulated using the software SRIM and results are shown as insets in Fig. S1a and Fig. S1b for the dose of $2\times10^{13}$ ions/cm$^2$ and $5\times10^{13}$ ions/cm$^2$, respectively. After annealing at 900 °C ions diffuse inside the Ge wafer. The Au ion distribution in Ge after diffusion is calculated and results are shown in Fig. S1a&b. The diffusion coefficient of Au in Ge at 900 °C is taken from Ref. [R1].

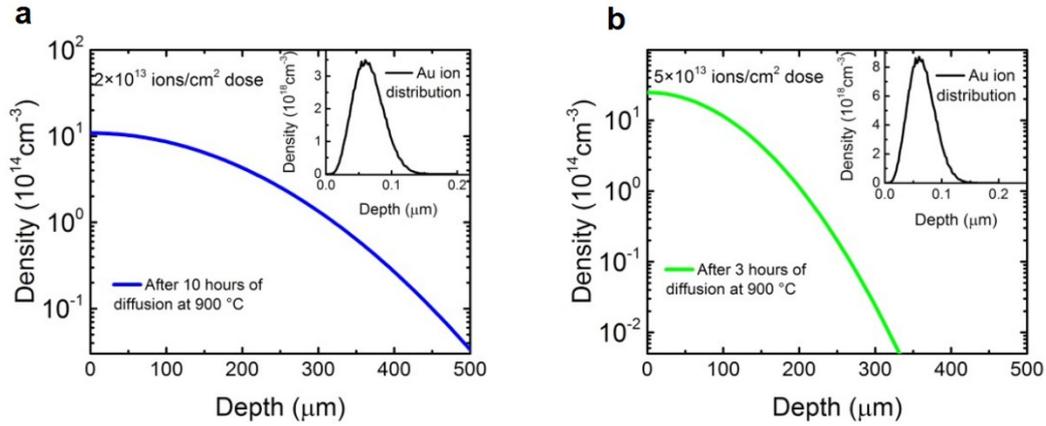

**Figure S1. Au concentration in Ge after ion irradiation and post anealing**. **a**, Ge:Au with the dose of 2x10$^{13}$ cm$^{-2}$ **b**, Ge:Au with the dose of 5x10$^{13}$ cm$^{-2}$. The insets show the distribution of the implanted Au ions before the annealing.

**Modelling of the photoinduced current**

In order to model the photocurrent we calculate the evolution of the near-infrared pump pulse during its propagation in the depth of the Ge:Au substrate. The temporal profile of the pump electric field at the given depth *z* is equal to

$$E(z,t)=\int_0^\infty E_0(\omega)e^{-\omega\kappa(\omega)z/c}e^{i\omega(n(\omega)z/c-t)}d\omega,$$

where $n^* = n(\omega) + i\kappa(\omega)$ is the complex refractive index of Ge, $E_0(\omega)$ is the spectrum of the pump pulse including its phase. In the numerical modelling, the integration is taken over the full bandwidth of the pump pulse.

The photogenerated carrier density is estimated as

$$N(z,\tau)=\frac{\alpha(\omega)\Phi(z,\tau)}{\hbar\omega}$$

where $\alpha$ is the absorption coefficient and $\Phi$ is the pump fluence. The simulation results for both pump wavelengths are shown in Fig. S2. The time scale is given in the lab reference system, i.e., it is corrected with respect to the propagation velocity of the THz pulse. Thus, the temporal spreading of the photoexcited carriers in the presented plots results solely from the difference between the group velocity of the pump pulse and the phase velocity of the THz transient. One can clearly see that in spite of the comparable duration for the pump pulses at 1100 and 1550 nm of 11 and 12 fs, respectively, the carrier excitation spreads over noticeably longer time in the latter case.

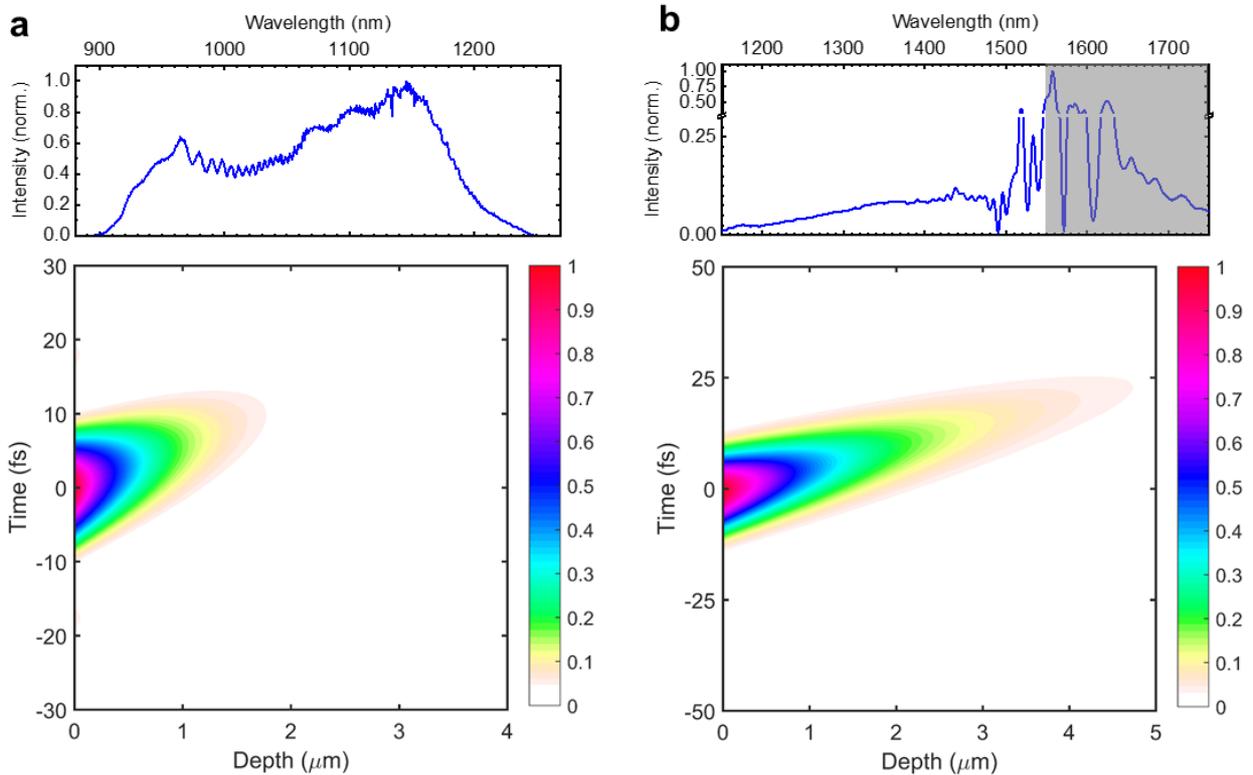

**Figure S2. Pump-induced carrier density $N(z,\tau)$ in the Ge:Au substrates as the function of the depth $z$ and the time $\tau$.** The time $\tau$ corresponds to the reference system of the electro-optic detector and taken into account the delay due to the propagation of the generated THz pulse through the substrate. **a**, Excitation at the center wavelength of 1100 nm; **b**, Excitation at the center wavelength of 1550 nm. Note different scaling in the graph. The upper panels show the respective spectra of the pump pulses and the shaded part in panel **b** marks the long wavelength part of the spectrum that does not contribute to the photoexcitation across the direct bandgap.

As discussed in the main text, this is related to the very weak absorption in Ge below the edge of the direct interband absorption that corresponds to ≈ 1550 nm (see Fig. 4a). Thus, only the short wavelength part of the spectrum contributes to the carrier generation resulting in the spectral narrowing and, consequently, in the temporal broadening of the photoinduced carrier distribution.

The transient photocurrent is calculated as a product of the carrier density and the bias electric field. Since the pump light penetrates for several micrometres into the substrate, it is necessary to know the spatial distribution of the applied dc electric field between the two electrodes of the emitter. We use COMSOL Multiphysics software to calculate the electric field distribution between the bowtie electrodes when a bias of 10 V is applied. The $y$-component of the electric field parallel to the THz polarization is shown in Fig. S3a for yz-plane passing through the centre of the bowtie structure. The electric field variation in the depth along the dotted line in the middle of the electrode gap (see Fig. S3a) is depicted in Fig S3b. The electric field drops rapidly up to 15 μm depth, but drops only slightly within the first 5 μm where most of charge carriers are generated. Definitely, the pumping at 1550 nm also generates carriers deeper in the substrate (beyond 5 μm depth) due to the absorption across the indirect bandgap, but according to our simulation their contribution to the photocurrent is negligibly small.

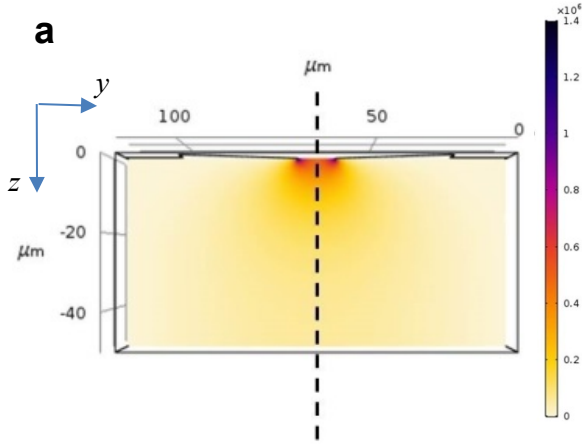

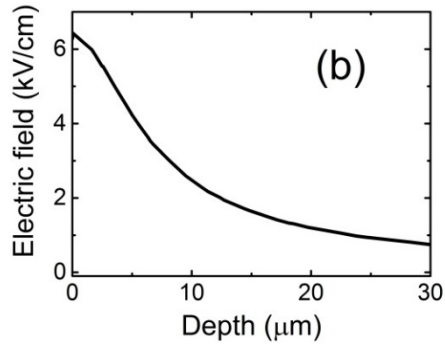

**Figure S3. Electric field distribution inside the Ge emitter. a**, y-component of electric field in a yz-plane passing through the center of the emitter; **b**, Electric field distribution along a line (dotted line in panel **a**) along the z-direction passing through the center of the emitter.

**Detector Response Function**

In order to calculate the detected THz field, the emitted THz radiation in the far field, which is proportional to the time derivative of the total photocurrent, has to be multiplied by the spectral response function of the electro-optic detector. The detector response functions of ZnTe detector crystals are calculated using the method given in reference.[30] The response function $R(\omega)$ is a function of THz frequency ($\omega$);

$$R(\omega) = G(\omega) \times r_e(\omega)$$

Where

$$G(\omega) = \frac{2}{n(\omega)+1} \times \frac{c \times [\exp\{-i2\pi\omega d(n_g(\lambda_0)-n(\omega))/c\}-1]}{-i2\pi\omega d(n_g(\lambda_0)-n(\omega))}$$

$$n(\omega)=\sqrt{\left[1+\{\frac{(\hbar\omega_{LO})^2-(\hbar\omega_t)^2}{(\hbar\omega_t)^2-(\hbar\omega)^2-i\hbar\gamma\omega}\}\right]\times\varepsilon_\infty}$$

and

$$r_{41}(\omega)=r_e\times\left[1+C\{1-\frac{(\hbar\omega)^2-i\hbar\gamma\omega}{(\hbar\omega_t)^2}\}^{-1}\right]$$

For ZnTe, $\hbar\omega_{LO}$ = 6.18 THz; $\hbar\omega_{TO}$ = 5.3 THz; γ= 0.09 THz; $\varepsilon_\infty$ = 6.7; C= -0.07 (according to Ref. 24); $r_e$ is a constant chosen as 1 here; and $n_g(\lambda_0)$ is the group refractive index of ZnTe at the probe pulse wavelength ($\lambda_0$). The probe pulse length and spectrum are shown in Fig S4a&b and Fig S5a&b for probe pulses used in setups with 1100 nm and 1550 nm pumping, respectively. Since the probe pulse is a broadband pulse and the group refractive index $n_g(\lambda_0)$ varies across the probe pulse spectrum, the detector response function $R(\omega)$ will also have different values for different probe wavelengths. The detector response function (DRF) is calculated at 13 discrete values of the probe wavelength (shown with square dots in Fig S4a and S5a. DRF at three such wavelengths are shown in Fig S4c and S5c by red, green and blue curves. DRFs calculated at all discrete wavelengths are then multiplied by the probe intensity at the corresponding wavelengths ($\lambda_0$) to get a weighted average of the DRF. Average DRFs are shown with black curves in Fig S4d and S5d.

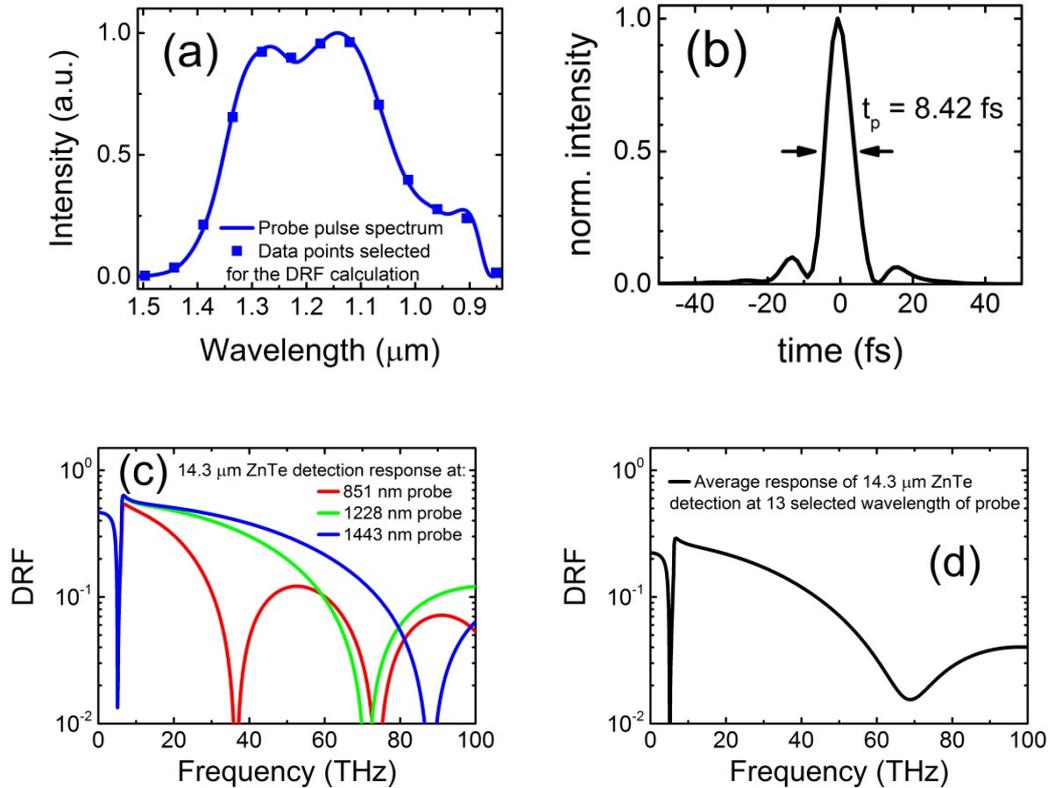

Figure S4. Detector response function for the setup with 1100 nm pumping. **a**, Spectrum of the probe pulse and chosen 13 data points for the calculation of DRF. **b**, Temporal profile of the probe pulse. **c**, DRFs at different probe wavelengths. **d**, Complete DRF obtained by weighted average of the DRFs across the probe spectrum.

**Frequency response due to pulse width used for electro-optic sampling**

The width of the pump pulse used to generate the charge carriers in the emitter is taken into account by our simulation of the transient photocurrent. On the other hand, the finite duration of the EOS probe pulse also limits the bandwidth of the recorded THz waveform. To determine the frequency roll-off due to the finite width of the probe pulse, the Fourier transform of the temporal pulse shape (shown in Fig. S4b and S5b) is multiplied with the DRF.

**Frequency response due to THz focusing on the detector**

At the detector crystal the NIR probe and THz pulse to be detected are focused as tightly as possible to get the maximum electro-optic signal. Since the focus spot size is limited by diffraction, smaller wavelengths are focused to smaller spot size. NIR probes have spot diameters much smaller than the THz spot diameter. The observed THz signal depends on THz electric field (square root of intensity) overlapping with the probe focus spot on the detector (actually overlapping volume in detector crystal). The THz intensity on the NIR probe focus spot is inversely proportional to the focus spot area of the THz wavelength; hence the observed THz signal is directly proportional to the THz frequency. This effect is included in the final simulation by multiplying the DRF with the corresponding THz frequency ($\omega$) and the Fourier transform of the probe pulse as mentioned above.

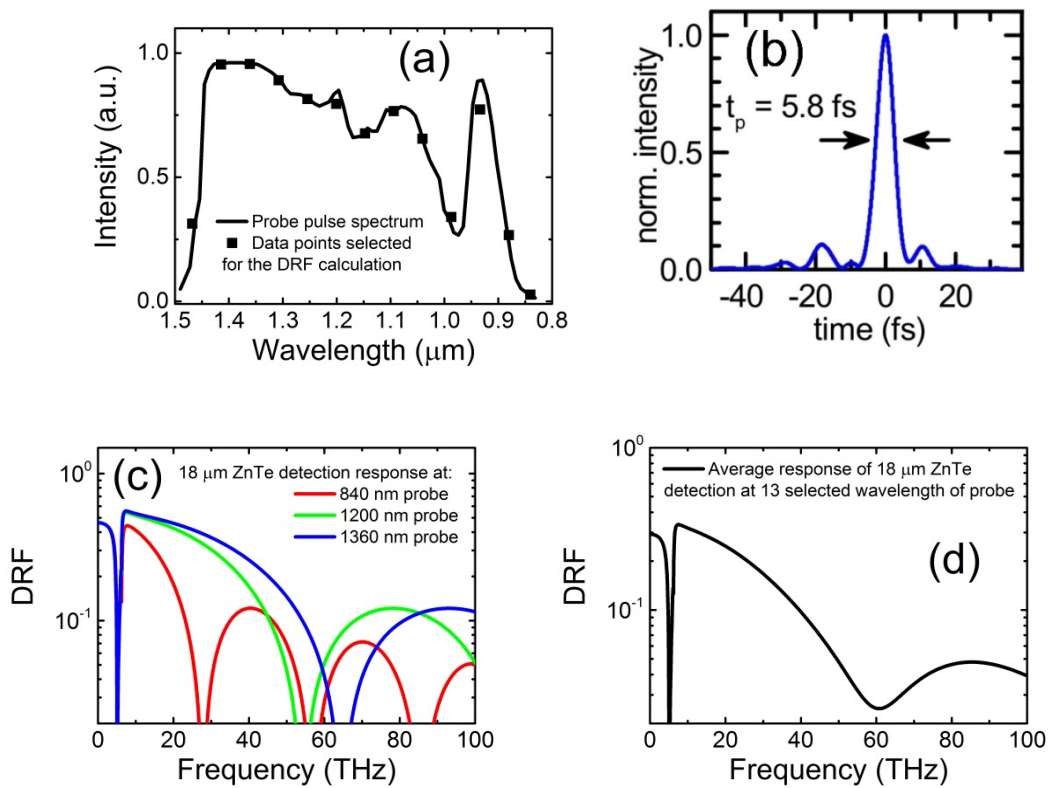

**Figure S5. Detector response function for the setup with 1550 nm pumping. a**, Spectrum of the probe pulse and chosen 13 data points for the calculation of DRF. **b**, Temporal profile of the probe pulse. **c**, DRFs at different probe wavelengths. **d**, Complete DRF obtained by weighted average of the DRFs across the probe spectrum.

**70 THz bandwidth with GaSe detector**

The THz signal emitted from Au implanted ($2\times10^{13}$ ions/cm$^2$) Ge based photoconductive THz emitter, when pumped with 1100 nm, is also detected with 18.4 µm GaSe electro optic crystal. Like the results shown in Fig. 4(b) in the main article using ZnTe detector, a THz spectrum up to 70 THz is recorded as shown in Fig. S6 (a&b).

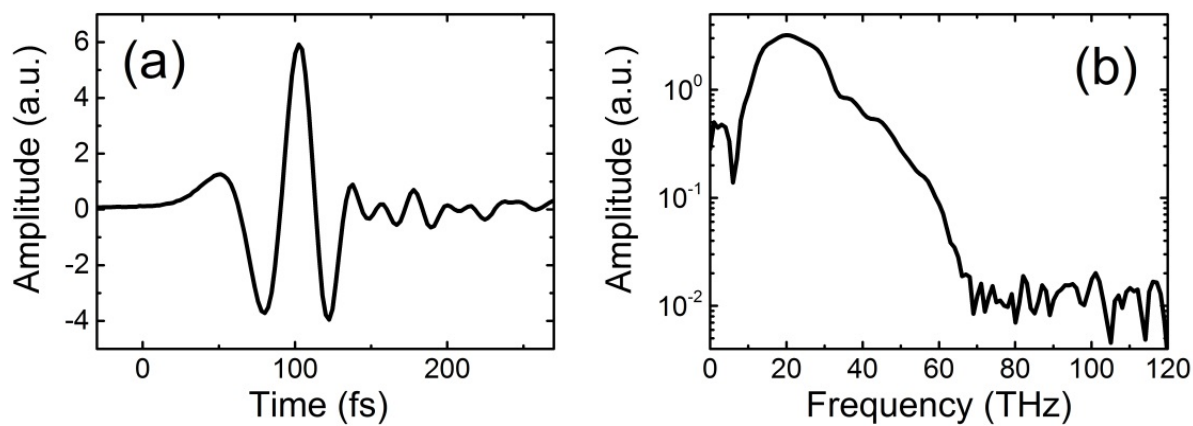

**Figure S6.** (a) THz pulse recorded using GaSe detector and (b) corresponding spectrum. Spectrum up to 70 THz is achieved.

**Emitter compatibility with 76 MHz repetition rate pumping**

To demonstrate the compatibility of implanted Ge emitters with a repetition rate even higher than 20 MHz, we tested the emitter with dose $5 \times 10^{13}$ ions/cm$^2$ using a conventional Ti:Sa oscillator operating at the wavelength of 800 nm and the repetition rate of 76 MHz. The pulse duration is nearly 100 fs and 1-mm-thick <110> ZnTe crystal is used as electro-optic detector. 100 mW pump power and 10 V (= 10 kV/cm electric field) bias are used on the emitter. The recorded time-domain pulse and its Fourier transform are shown in Fig. S7. The high-frequency cut off is below 4 THz due to the thickness of the ZnTe detector and the laser pulse width used for generation and detection of the THz pulses. However, the performance of the Ge:Au antenna is comparable to its GaAs-based analogue in agreement with our previous study.[22]

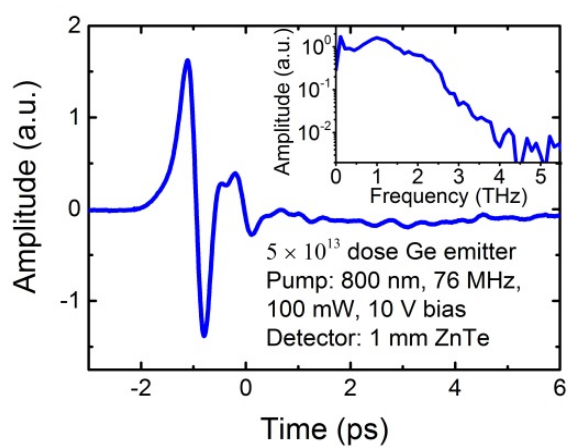

**Figure S7. Emitter performance with 76 MHz, 100 fs, 800 nm.** THz pulse emitted from implanted Ge emitters when pumped with ~ 100 fs, 800 nm Ti:Sa laser pulses at 76 MHz. The spectral bandwidth is limited due to the pulse width of pump and probe pulses, and the thickness of the ZnTe detector crystal.